\documentclass[12pt]{article}

\newcommand{\comment}[1]{}
\newcommand{\keywords}[1]{}

\begin{document}

\title{Coordinate System, Temperature and Gravity}
	\author{Victor Olkhov \\
	Institute for High Temperatures\\
	Lab.89, Krasnokazarmenaia 17-a\\
	Moscow, 111250, Russia\\
	email:ipvicol@mail.cityline.ru}
\date{}
\maketitle

\begin{abstract}
We discuss the problem of applicability of Coordinate Systems (or Frames)
that determine $(t,x,y,z)$ values - the initial notions for most physical
theories. Equipment that measure these values - Clocks and Meters - are
based at Reference System and are the primary measuring units. We discus
when and why physical phenomena might prevent to provide measurements of 
$(t,x,y,z)$. We show that Temperature may be the factor that can
significantly influence on the measurements of $(t,x,y,z)$ by Reference
System and that action may violate the usage of Coordinate System. We
discuss possible origin of such unmovable Temperature and assume that it
should be Gravity.
\end{abstract}

\newpage

\section{Introduction}

\label{section:intro} Thermodynamical and Chaotic behavior in the presence
of strong Gravity were studied a lot during last 30 years (\cite{wa}, \cite
{ma}, \cite{gr} and ref. there in). In this notes we would like to attract
ones attention to small aspect of this advanced matter. All notions and
models that are used in Physics have their own applicability conditions. For
example the usage of geometric optics approximation, continuum media
approximation, free harmonic oscillator etc. are reduced by certain physical
factors. To check up the usage conditions one make experiment to compare
it's results with applicability conditions and predictions of the model.
Having that in mind we are interested in the problem of possible reduction
of usage of Coordinate System ($CS$) or Frame - fundamental notion that is
in the basement of Physical models and theories. The problem we are going to
discuss concern possible reduction of usage of $CS$ due to Thermodynamical
factors and hence relations between Temperature and Gravity are of high
interest.

\section{Coordinate system usage conditions.}

\label{csuc} $CS$ - is a mathematical notion, as well as Observer, Frame
etc., used to determine $(t,x,y,z)$ coordinates and metrics and is the
basement for geometrical approach to Gravity and Space-Time description. All
physical processes are described in certain $CS$ and $(t,x,y,z)$ variables
are the most fundamental physical values. The applicability (or non
applicability) of $CS$ for the description of the given physical state is
determined by the chance (or no chance) for that particular physical state
in real (or imaginary) experiment to provide the measurements of $(t,x,y,z)$
by physical equipment. Such measurements in real or imaginary experiment are
realized by special unit - Reference System $(RS)$. We treat $RS$ as a
physical equipment, like Meter and Clocks, to measure main physical
properties - Time and Distance $(t,x,y,z)$. As well $RS$ has equipment for
transmitting and receiving measured information from and to other $RS$.

It is well known that $(t,x,y,z)$ coordinates determined by certain $CS$ are
the only parameters that characterize local geometry of Space-Time. $CS$
determine $(t,x,y,z)$ free from other physical properties, hence $RS$
measuring unit that confirm the usage of given $CS$, must to measure $%
(t,x,y,z)$ free from any external physical influence. Experiments prove that
one able to reduce and avoid the influence of electromagnetic fields,
particles, thermal influence of surrounding media, rain, wind, etc. on the
measurements of $(t,x,y,z)$. Ability to avoid any external influence on
measurements of $(t,x,y,z)$ equals applicability of $CS$ and physical models
based on that notion.

Let's assume now, that for certain physical state it is actually impossible
to avoid or to reduce the influence of surrounding physical factors,
temperature in particular, on the measurement units of the $RS$. Suppose,
that such unmoved Thermal action significantly act on the measurements of 
$(t,x,y,z)$ and information exchange. Imagine, that Temperature $T$ of the 
$RS$ units for such physical state, as well as temperature of any other bodies
or electromagnetic fields equals $10^5K$ and can not be decreased by any
means. Suppose, that no screens, coolers or any other physical device or
effects exist, that can protect the measuring devices from such temperature
or to reduce it value. Then, even if one assume that certain Meter and
Clock, as well as receiver/transmitter of information can ever exist at such
high temperature, the results of the measurements for such exotic state must
depend on the value of the temperature. Then such unmovable non-reduced
temperature $T$ should be treated as fundamental property of such exotic
state, as $(t,x,y,z)$. Appearance of additional physical factor -
Temperature - at least as fundamental as $(t,x,y,z)$ - in the set of primary
observables quit applicability of common $CS$ and physical models based on
it for the description of that particular physical state. Even if $(t,x,y,z)$
do not depend on the value of Temperature directly, one should find new
fundamentals to describe physical state with primary observables $(T,t,x,y,z)
$, where $T$ is the min Temperature obtained for given $CS$ .

Why it is difficult to describe such exotic physical state with primary
measurable values $(T,t,x,y,z)$ where temperature $T$ define unmovable non
reduced temperature of the local area measured by $RS$? We will mention only
few problems.

\begin{itemize}
\item  Since Temperature become fundamental property, like Time, than
Thermodynamics must be treated as fundamental theory, and it can not be
derived from Mechanics or other theories, based on $CS$. Statistical
ensembles and other notions of Statistical Physics have no usage for this
case.

\item  Most Mechanical notions lose applicability and one not able to use
Free Space approximation, Vacuum states, Phase Space, Kinetic description
and other common models. $CS$ has no use for such states.

\item  Unmoved Thermal action on the instruments means that all classical
measurements have finite accuracy and thus no small Time or Distance scales
exist in such physical state.
\end{itemize}

\section{Thermal Gravity}

\label{tg} Do such states exist? What can be the origin of such unmoved and
non reduced temperature and why it might be impossible to protect the
instruments from the temperature action?

As usual Thermal properties are associated with Matter and Electromagnetic
fields (mainly). Temperature of bodies, temperature of particles,
temperature of continuum media, temperature of radiation - are the main
sources of Thermal influence. Present experience prove that one able to
reduce such Thermal action and to protect $RS$ measuring units from such
influence.

We suppose that Gravity itself should be origin and source of unmovable
Thermal influence for certain physical states. At the same time we remind
internal contradiction of our treatment: properties of present physical
models are based on the applicability of $CS$. We use them to demonstrate
possible existence of Thermal properties of Gravity. That violate the use of 
$CS$ and models based on it. So, our reasons show definite internal
contradiction of modern physical fundamentals, at least.

The physics of Thermal Gravity component assumed similar the Electromagnetic
Thermal fluctuating radiation \cite{ru}: hot matter radiate Thermal
Electromagnetic field and this Thermal fluctuating field heat matter.

Similar to that \cite{ol} Free Fall Acceleration $(FFA)$ induced by any
macro body consist of the main averaged part (the main factor) and
negligible fluctuating Thermal component that reflect the Thermal state of
the particles of the body. Thermal component of $FFA$ is neglected due to
it's careless vanishing influence but it is not correct to forget it at all.
(We use $FFA$ as simplest property of Gravity).

The significant property of this small Thermal constituent of $FFA$ is
ability by direct action to heat any matter and Electromagnetic field.

Otherwise it will contradict with the Second Law of Thermodynamics 
\cite{hu}. If Thermal $FFA$ component induced by Thermal behavior 
of particles of macro body will produce only regular movement of test 
particles without any heating, then one will have the way of direct 
transfer of any value of heat into regular mechanical movement. 
Such process should decrease the Entropy for sure.

It is lucky chance the coupling $const$ of Gravity is much more less than
for Electromagnetic interaction. Due to that Thermal component of Gravity is
careless and $RS$ measurements of $(t,x,y,z)$ describe reality with high
accuracy. But size of coupling $const$ can not forbid existence of that
effect.

The most significant is the actual existence of Thermal Gravity component.
Let us mention Bondi \cite{bo}, who in 1961 pointed out possible existence of
gravity fields induced by atoms and even by protons and electrons. But, due
to Uncertainty Principle he assumed that it is impossible to measure these
values. We suppose, that universal Thermal action of such Gravity fields,
and those induced by molecules, is exactly the factor that prevent such
precise measurements. One able to measure only Thermal action of such
Gravity, induced by warm particles, and not able to measure these values in
pure dynamical sense. It seems that such Thermal reduction of Gravity
measurements accuracy is much more rough than $h/2$ factor and that also
should be the trouble on the way as to Quantum Gravity as to Gravity 
waves detection.

In the ''area'' with significant value of Thermal $FFA$ component all
bodies, fields, equipment of $RS$ will be heated up to certain value of
Temperature $T_g$. That temperature will be unmovable and no physical means
exist that can decrees this Temperature due to universal action of Gravity.
One able to protect the $RS$ units from Thermal electromagnetic influence,
but fail to reduce or reflect the action of Thermal $FFA$ heating effect.
Anything that interact with Gravity will be heated by it's Thermal
component. The Temperature $T_g$, induced by such Thermal Gravity component
become the fundamental property of the area and must be treated as basic as 
$(t,x,y,z)$ if they can be ever measured. Indeed, Temperature of instruments
will reduce accuracy of measurements. On the other hand high Temperature
Gravity may thermalize any regular signal and the only information that will
be received by external $RS$ will be only Thermal. In such area the set 
$(T_g,t,x,y,z)$ should be treated as primary measurable values and that
violate usage of $CS$.

And what can be said about the High Temperature Gravity areas? Do they exist?

We assume, that High Temperature Gravity should exist and should be measured
by $RS$ near the Infinite Red Shift surface. Thermal equilibrium conditions 
\cite{la} requires $(g_{_{00}})^{1/2}=const$. As $g_{_{00}}$ vanish to zero
at infinite red shift surface (Shwarzshild radius $r_g=2GM/c^2$ for mass 
$M$) then $T_g\rightarrow \infty $ if it is positive far from Black Hole. That
violate the usage of $CS$ (or ''Observer'' notion) for certain $r_{cr}>r_g$
due to critical value of Temperature $T_{cr}$.

The second case of high temperature action of Gravity should exist for the 
$RS$ that moves with near light speed with respect to certain preferred 
$RS_p$. Indeed, if small Temperature of Gravity $T_g$ exist than one can 
chose $RS_p$ that measures the lowest value of such Temperature. 
The Observer in the $RS$ that moves with speed $v$ with respect to 
$RS_p$ should measure \cite{ha} the Temperature of Gravity 
$T=T_g(1-b^2)^{-1/2}$, $b=v/c$. Thus $RS$ will be under critical Thermal 
action of Gravity if it moves with critical speed $v_{cr}<c$ with respect 
to preferred $RS_p$. That means that Thermodynamics have it's own 
restrictions for speed $v<v_cr<c$ to reach the speed of light value 
and that reduce applicability for $CS$ as well.

Our considerations on possible violation of applicability of $CS$ do not
touch the usage of most physical phenomena, except Black Holes physics and
description of Universe as a whole. Possible existence of self Gravity
Temperature violate usage of $CS$ {\it above} infinite red shift surface
and reduce acceptable speed value by certain $v_{cr}<c$. That means that
Gravity can not be described by models based on $CS$ anywhere and Universe
evolution models might be restudied. We suppose that one else explanation of
observed Microwave Background Radiation can be discussed: this radiation
should reflect averaged Temperature of Gravity in our area and it's
Temperature should be treated as Gravity's Temperature $T_g=2.7K.$

Conditions for which one should take into account the Temperature of Gravity
and will fail to use $CS$ as basement of the theory are more than exotic.
For example, to measure $T_g=1000K$ it is required $\Delta
r=r-r_g=r_g(T_g/2.7)^{-2}=r_g10^{-6}$ ; $\Delta
v=c-v=c(T_g/2.7)^{-2}=10^4cm/c$

We are lucky to live in the cold Gravity area and thus able to use $CS$ to
describe Physics. If Temperature of Gravity really exist, then Gravity
itself should be treated as origin of Thermal properties. For certain exotic
cases Thermal and Space-Time variables can not be measured separately. Some
new approach to describe such states should be suggested. In any case the
relations between Gravity and Temperature and their measurements problem
should be studied.

\section{Conclusions. }

\label{con} A lot of {\it Pro} and {\it Contra} arguments may be added
to the reasons that we briefly presented above to demonstrate possible
existence of self Gravity Temperature, corresponding measurements problem
and applicability of Coordinate System problem. But the questions: What are
actually the applicability conditions of Coordinate System? Can
Thermodynamics and Temperature in particular make any difficulties for
measurements procedures? Do Gravity have Temperature and how it can be
measured? - these and some other questions should be studied to make our
understanding of the relations between Thermodynamics and Gravity more
precise.


\begin{thebibliography}{99}

\bibitem{wa}  R.M. Wald, ``The Thermodynamics of Black Holes'',
gr-qr/9912119,

\bibitem{ma}  E.A.Martinez, {\it The postulates of Gravitational
thermodynamics}, Phys. Rev. {\bf D54}, 6302 -6346 (1996), gr-qg/9609048

\bibitem{gr}  Proc. Eight Marsel Grossmann Meeting on General Relativity,
Ed. T.Piran, R.Ruffini, W.Science Publ.,(1999), Sec.:Chaos in General
Relativity, 612-641, Sec.:Black Holes thermodynamics, 959-1002, G.Hurwitz,
``Thermodynamics for Gravitating Systems is Different'', 846-861

\bibitem{ru}  M.L.Levin, S.M.Rutov, ``Theory of Equilibrium Thermal
Fluctuations in Electrodynamics'', Moscow, Science Publ., (1967).

\bibitem{ol}  V.M.Olkhov, ``Physical reasons that reduce applicability of
Relativity'', Late Papers Proc., Physical Interpretation of Relativity
Theory IV, Ed. M.C.Duffy, London, 177-183 (1998)

\bibitem{hu}  K.Huang, ``Statistical Physics'', J.Wiley Inc.,N.Y. (1963)

\bibitem{bo}  H.Bondi ``Gravitational waves'', Endeavour, {\bf XX}, {\bf 79},
121-130 (1961)

\bibitem{la}  L.D.Landau, E.M.Lifshitz, ``The Field Theory'', Moscow,
Science Publ., (1967)

\bibitem{ha}  D.ter Haar, G.Vergeland, ``Thermodynamics and Statistical
Mechanics in Special Relativity'', in Einstein collection, {\bf 72}, Moscow,
Science Publ., p.254, (1974)


\end{thebibliography}
\end{document}